# Evidence-Based Robust Design of Deflection Actions for Near Earth Objects

Federico Zuiani
*PhD Candidate,*
School of Engineering, University of Glasgow, James Watt South Building, Glasgow, G12 8QQ, United Kingdom
Tel.: +44-141-5484558
f.zuiani.1@research.gla.ac.uk

Massimiliano Vasile
*Reader,*
Department of Mechanical & Aerospace Engineering, University of Strathclyde, 75 Montrose Street, Glasgow, G1 1XJ, United Kingdom
Tel.: +44-141-3306465
Fax: +44-141-3305560
massimiliano.vasile@strath.ac.uk

Alison Gibbings
*PhD Candidate,*
School of Engineering, University of Glasgow, James Watt South Building, Glasgow, G12 8QQ, United Kingdom
Tel.: +44-141-5484558
a.gibbings.1@research.gla.ac.uk

**Abstract.** This paper presents a novel approach to the robust design of deflection actions for Near Earth Objects (NEO). In particular, the case of deflection by means of Solar-pumped Laser ablation is studied here in detail. The basic idea behind Laser ablation is that of inducing a sublimation of the NEO surface, which produces a low thrust thereby slowly deviating the asteroid from its initial Earth threatening trajectory. This work investigates the integrated design of the Space-based Laser system and the deflection action generated by laser ablation under uncertainty. The integrated design is formulated as a multi-objective optimisation problem in which the deviation is maximised and the total system mass is minimised. Both the model for the estimation of the thrust produced by surface laser ablation and the spacecraft system model are assumed to be affected by epistemic uncertainties (partial or complete lack of knowledge). Evidence Theory is used to quantify these uncertainties and introduce them in the optimisation process. The propagation of the trajectory of the NEO under the laser-ablation action is performed with a novel approach based on an approximated analytical solution of Gauss' Variational Equations. An example of design of the deflection of asteroid Apophis with a swarm of spacecraft is presented.

# 1. Introduction

During the last two decades, Near Earth Objects (NEO) have attracted considerable interest from the scientific community in general and in particular in the space field. The reasons for this are twofold: first, from a strictly scientific point of view, asteroids can provide precious data to reconstruct the genesis of the solar system. In this sense, NEOs, in contrast to other small celestial bodies, are relatively easy to reach and explore, thanks to their small dimensions, lack of atmosphere and vicinity to the Earth. On the exploration side, there is a number of past or ongoing missions aimed at the study of small celestial bodies, such as NEAR (McAdams et al. 2000), Rosetta (Glassmeier et al., 2007), Deep Space 1 (Rayman et al. 2000), Hayabusa (Nakamura and Michel 2009), Deep Impact (Hampton et al. 2005) and Dawn (Russel et al. 2007).

The second reason instead is linked with the potential threat they represent for our planet. According to the most recent tracking data, over 1000 NEOs have been classified as potentially hazardous to the Earth, i.e. they have an Earth Minimum Orbit Intersection Distance (MOID) of 0.05 AU or less and an absolute magnitude of 22.0 or less (JPL 2012). This suggests that the danger of a catastrophic event in the mid to long term is not unrealistic. The historical perspective of past impact events (e.g. Tunguska in 1908) is an important reminder of the dire consequences this could have on our fragile ecosystem.

Therefore, the scientific community has proposed a number of mitigation strategies and techniques to counteract the hazard of a NEO impact. The first serious technical study, Project Icarus (MIT 1968), dates back to 1967 but only in the 90s the theme has started to be widely explored by scientists and engineers and various strategies have been proposed. Among them we find techniques producing an impulsive change in the asteroid motion such as Nuclear blast (Smith, Barrera et al. 2004) and Kinetic Impactor (McInnes 2004), or attached Chemical engines (Scheeres and Schweickart 2004); there are others which produce a continuous low thrust like in the case of using attached Electrical thrusters (Scheeres and Schweickart 2004), or electrically propelled gravitational tugs (Lu and Love 2005), or by means of the low thrust produced by surface Ablation, the latter induced either by solar collectors (Melosh and Nemchinov 1994) or laser beam (Campbell, Phipps et al. 2003). Other more exotic systems include Mass Drivers (Olds, Charania and Schaffer 2007), which involve the controlled ejection of asteroid's surface material in order to produce a series of small impulsive changes in its motion; there are proposals also for passive methods, like the idea of painting part of the asteroid to modify its optical properties and thus take advantage of the Yarkovsky effect (Spitale 2002).

A recent study (Sanchez, Colombo et al. 2009) presented a quantitative comparison of different deflection methodologies that suggested that surface ablation techniques could represent an advantage compared to other methodologies.

The principle behind the surface ablation strategies is that of inducing the sublimation of the surface material of the asteroid. This will create an ejecta plume and an associate small continuous thrust. This thrust, over extended periods of time, will slowly deviate the asteroids from its initial orbit. Ablation strategies based on direct irradiation with concentrated solar light were proposed by Melosh and Nemchinov (1994) who envisioned using a single large solar concentrator to irradiate a relatively small spot on the surface of the asteroid so that the resulting heat will induce the sublimation. Other authors proposed the use of lasers in conjunction with a nuclear power source (Phipps, 1992, 1997, 2010; Park and Mazanek, 2005). Extensive studies on the dynamics of the deflection with high power lasers were proposed by Park and Mazanek (2005) envisaging a single spacecraft with a MW laser. The combination of solar concentrators with lasers (directly or indirectly pumped) was recently proposed by Maddock and Vasile in 2008. The idea is to use a formation of smaller concentrators, each powering a solar-pumped laser. Thus, the spacecraft could be placed further from the NEOs, in this way also avoiding almost entirely the contamination due to the ejecta plume. Recent numerical and experimental analyses (Vasile et al. 2009a, 2009b; Maddock and Vasile 2008; Gibbings, Vasile et al. 2011a, 2011b) have already investigated the basics of the solar-pumped, laser ablation concept. There are, however, some epistemic uncertainties on the physical properties of the asteroid and on some design low Technology Readiness Level (TRL) components of the spacecraft. This work addresses the impact of uncertainties on the performance of the laser system. In order to do so, an approach based on Evidence Theory is introduced (Vasile and Croisard 2010). This approach requires the evaluation of several deflected asteroid trajectories. The computation of the deflected trajectory under the effect of laser ablation is here performed with a novel approach based on an approximated solution of Gauss' planetary equations (Zuiani, Vasile et al. 2011). The paper is organised as follows: after introducing the models for the trajectory, the spacecraft system and the deflection action, the uncertainties are analysed and quantified through Evidence Theory. A multi-objective optimisation problem is then solved to find optimal deflection solutions under uncertainty. The paper then presents an analysis of sensitivity to identify which epistemic uncertainty is the most significant in the context of asteroid deflection with laser ablation.

## 2. Trajectory and Deflection Model

In order to assess the performance of the laser ablation approach, a hypothetical asteroid based on 99942 Apophis is considered. Its orbital elements are suitably modified in order for it to intercept the Earth in 2036. The effectiveness of the deflection action is measured by the magnitude of the impact parameter *b* with respect to the Earth at the time of the expected collision, as shown in Fig. 1 where $V_\infty$ is the incoming velocity of the asteroid and $\mathbf{v}_E$ is the velocity of the Earth. The impact parameter is computed by projecting the deviated position of the asteroid on the Earth's b-plane at the epoch of the expected impact (Vasile and Colombo 2008). In this case study, the undeviated orbit has *b*=0.

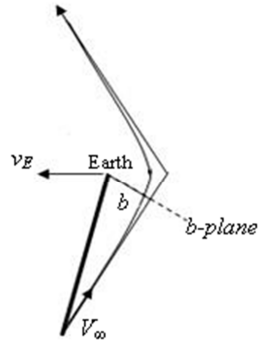

**Fig. 1** Impact parameter

The computation of *b* requires the variation of the orbital elements due to the deflection action. From the variation of the orbital elements one can use the deflection formulas in (Colombo et al. 2009) or the nonlinear proximal motion equations in (Vasile and Maddock 2010) to compute the position and velocity relative to the Earth. The variation of the orbital elements is obtained by integrating Gauss' Variational Equations in non-singular equinoctial elements, (Battin 1999):

$$\frac{da}{dt} = \frac{2a^2}{h}\left[(P_2 \sin L - P_1 \cos L)\varepsilon \cos\beta \cos\alpha + \frac{p}{r}\varepsilon \cos\beta \sin\alpha\right]$$

$$\frac{dP_1}{dt} = \frac{r}{h}\left\{-\frac{p}{r}\cos L \cdot \varepsilon \cos\beta \cos\alpha + \left[P_1 + \left(1+\frac{p}{r}\right)\sin L\right]\varepsilon \cos\beta \sin\alpha - P_2(Q_1 \cos L - Q_2 \sin L)\varepsilon \sin\beta\right\}$$

$$\frac{dP_1}{dt} = \frac{r}{h}\left\{-\frac{p}{r}\cos L \cdot \varepsilon \cos\beta \cos\alpha + \left[P_2 + \left(1+\frac{p}{r}\right)\sin L\right]\varepsilon \cos\beta \sin\alpha - P_1(Q_1 \cos L - Q_2 \sin L)\varepsilon \sin\beta\right\} \quad (1)$$

$$\frac{dQ_1}{dt} = \frac{r}{2h}\left(1+Q_1^2+Q_2^2\right)\sin L \cdot \varepsilon \sin\beta$$

$$\frac{dQ_2}{dt} = \frac{r}{2h}\left(1+Q_1^2+Q_2^2\right)\cos L \cdot \varepsilon \sin\beta$$

$$\frac{dL}{dt} = \sqrt{\frac{\mu}{a^3}} - \frac{r}{h}(Q_1 \cos L - Q_2 \sin L)\varepsilon \sin\beta$$

where the six non-singular equinoctial elements are defined as:

$$\begin{aligned}
P_1 &= e\sin(\Omega+\omega) \\
P_2 &= e\cos(\Omega+\omega) \\
Q_1 &= \tan\frac{i}{2}\sin\Omega \\
Q_2 &= \tan\frac{i}{2}\cos\Omega \\
L &= \Omega+\omega+\theta
\end{aligned} \quad (2)$$

and $a$ is the semi-major axis, $e$ the eccentricity, $i$ the inclination, $\Omega$ the right ascension of the ascending node (RAAN), $\omega$ the argument of periapsis, $\theta$ the true anomaly, $r$ is the radius, $h$ the angular momentum, $p$ the semi-latus rectum, $\mu$ the gravity constant of the central body and $L$ the true longitude. $\varepsilon$, $\alpha$ and $\beta$ are respectively the modulus, azimuth and elevation of the thrust acceleration in the radial-transversal reference frame, as in Fig. 2, forming the thrust vector:

$$\mathbf{f} = \varepsilon[\cos\alpha\cos\beta \quad \sin\alpha\cos\beta \quad \sin\beta]^T \quad (3)$$

Numerical integration of Gauss' Variational equations would be too computationally expensive for the analyses in this paper, as thousands of trajectories need to be evaluated. Hence, Gauss' equations are here integrated using Finite Perturbative Elements in Time (FPET). FPET are based on an approximated analytical solution of Gauss' equations over short arcs for a constant thrust (Palmas 2010; Zuiani, Vasile et al. 2011). The next section will describe the derivation of the FPET approach.

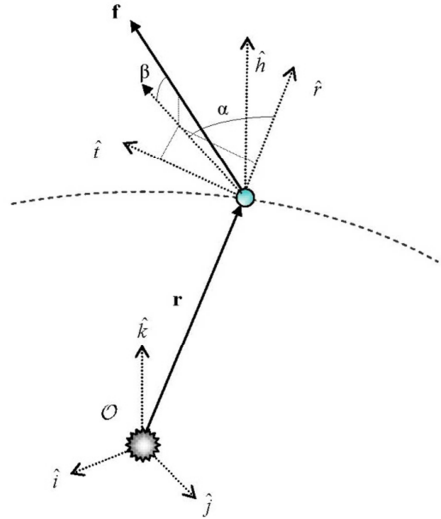

**Fig. 2** Radial-transversal reference frame

### 2.1 The Low Thrust Perturbative Approach

Assuming that $\varepsilon$ is small compared to the local gravitational acceleration (the ablation induced acceleration is in the range $10^{-7}$-$10^{-12}$ m/s$^2$) and that the thrust modulus and direction are constant in the radial transversal reference frame over an arc of length $\Delta L$, one can expand the orbital elements and time up to the first order in the perturbing parameter $\varepsilon$ as follows:

$$a(L) = a_0(L_0) + \varepsilon\, a_1(\Delta L, \alpha, \beta)$$
$$P_1(L) = P_{10}(L_0) + \varepsilon\, P_{11}(\Delta L, \alpha, \beta)$$
$$P_2(L) = P_{20}(L_0) + \varepsilon\, P_{21}(\Delta L, \alpha, \beta)$$
$$Q_1(L) = Q_{10}(L_0) + \varepsilon\, Q_{11}(\Delta L, \alpha, \beta)$$
$$Q_2(L) = Q_{20}(L_0) + \varepsilon\, Q_{21}(\Delta L, \alpha, \beta)$$
$$t(L) = t_{00}(L_0, \Delta L) + \varepsilon\, t_{11}(\Delta L, \alpha, \beta)$$

(4)

where:

$$L = L_0 + \Delta L \tag{5}$$

The zero-order terms, obtained for $\varepsilon = 0$ correspond to the unperturbed Keplerian motion. Once the analytical expressions for $a_1(L)$, $P_{11}(L)$, $P_{21}(L)$, $Q_{11}(L)$, and $Q_{21}(L)$ are available, together with $t_{00}(L)$ and $t_{11}(L)$, the variations of the five orbital parameters and time are known as a first order approximated function of the true longitude $L$, with respect to the reference state at $L_0$. Thus, one can analytically propagate the non-singular elements, either backward or forward in $L$, for an arbitrary set of initial (final) conditions and control force components, expressed in terms of magnitude and two angles (Palmas 2010).

Motion propagation is obtained by subdividing the trajectory into Finite Perturbative Elements in Time, i.e. a number of arcs each characterised by a constant thrust acceleration vector in the radial transversal reference frame (as shown in Fig. 3). Within each arc, forward propagation of the motion is performed analytically.

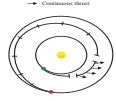

**Fig. 3** Low Thrust trajectory subdivided in FPET arcs

In a previous work (Zuiani, Vasile et al. 2011) it was shown that this analytical propagation with FPET allows for computational times of at least one order of magnitude lower than numerical integration but with comparable accuracy. In this paper, at the beginning of each trajectory arc, the thrust acceleration acting on the NEO is computed by evaluating the ablation model (see Section 4.). Since, in general, the thrust magnitude varies with a periodic pattern along the trajectory, the ablation model needs to be evaluated quite often. The frequency with which the model is evaluated dictates the amplitude of the trajectory arcs. The basic idea is to have short arcs when the thrust is high and larger ones when the thrust is low. In order to achieve this, during the propagation the arc length $\Delta L$ is dynamically adjusted with the simple law:

$$\Delta L = \min\left[ A \exp\left(\frac{-\log_{10}\varepsilon + \log_{10}\varepsilon_{max} + 1}{k}\right)\ \ \Delta L_{max}\right] \tag{6}$$

where $\varepsilon$ is the current value of the thrust acceleration, $\varepsilon_{max}$ is the largest value it has assumed so far and $A$, $k$ and $\Delta L_{max}$ are constants which were tuned empirically in order to achieve a good compromise between accuracy and CPU cost compared to the numerical integration. This was done by performing a high number of propagations of the trajectory and ablation models with different candidate sets of tuning parameters. As a result, the set which guaranteed a negligible error on the impact parameter $b$ with respect to the numerical integration at the lowest computational cost was chosen. As an example, using FPET to propagate the trajectory and ablation models implemented in Matlab® on a Intel Core Duo® 3.16 GHz machine running Windows 7® e requires between 0.2 and 2 seconds (depending on the length of the trajectory), compared with up to 30 seconds when using numerical propagation.

### 3. Spacecraft System Model

The solar-pumped laser ablation concept envisions the use of a formation of $n_{sc}$ identical spacecraft, each provided with a solar-pumped laser system. These will be flying in the proximity of the asteroid (see Fig. 4) with a distance from the asteroid's surface between 1 and 4 km (Vasile and Maddock 2008). Note that the plume shape in Fig. 4 is a qualitative depiction of the contamination model by Kahle et al. (2006) as in Section 4.

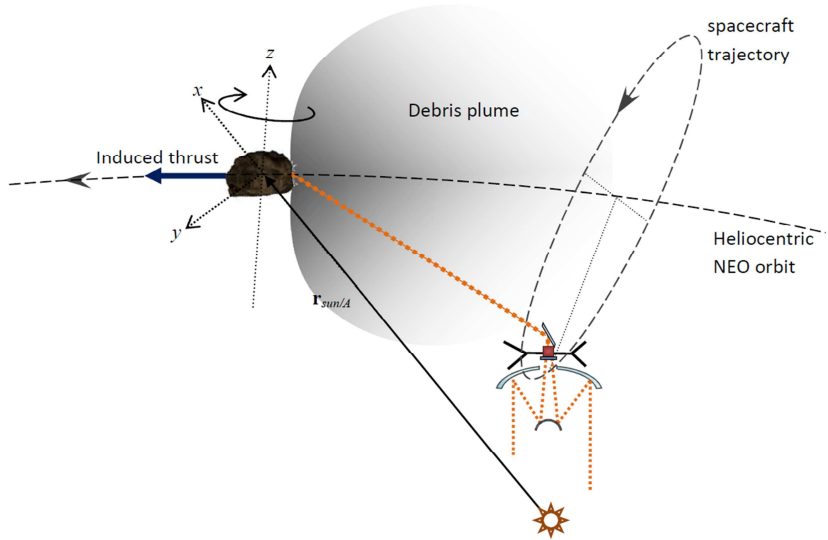

**Fig. 4** Spacecraft's proximal motion with respect to the asteroid

Each spacecraft in the formation (see Fig. 5) is composed of a large primary mirror $M_1$, which focuses the solar rays on a smaller secondary mirror $M_2$. The solar rays are then conveyed onto a solar array $S$, which powers a laser plus other subsystems. The laser beam is directed towards the NEO by means of a directional mirror $M_d$. A set of radiators dissipates the excess heat in order to keep the temperature of the solar array and the laser within operational limits.

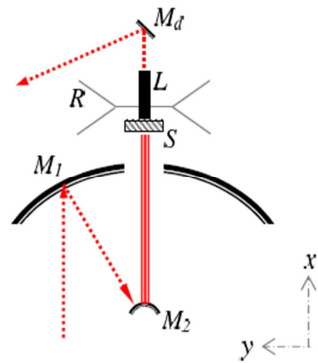

**Fig. 5** Laser spacecraft system

The dry mass of the spacecraft is computed as:

$$m_{dry} = k_{dry}\left(m_C + m_S + m_M + m_L + m_R + m_{bus}\right) \qquad (7)$$

where $m_C$ is the mass of the harness, $m_S$ is the mass of the solar arrays, $m_M$ is the mass of the mirrors, $m_L$ is the laser mass, $m_R$ is the radiator mass and $m_{bus}$ is the mass of the bus and the constant $k_{dry}$ represents the margin on dry mass. The masses of the various subsystems are computed with simple analytical formulas. The harness mass is expressed as a fraction of the combined mass of the laser and solar array:

$$m_C = MF_C (m_S + m_L) \qquad (8)$$

The radiator mass $A_R$ is proportional to the area needed to dissipate the excess power. $MF_C$ is the mass fraction for harness. The latter is computed from a steady state thermal balance between the Solar input power and the emitted power which is not reported here for the sake of conciseness.

$$m_R = \rho_R A_R \qquad (9)$$

where $\rho_R$ is the radiator specific mass per surface unit area. The mass of the solar arrays is proportional to their area $A_S$:

$$m_S = k_S \rho_S A_S \qquad (10)$$

where $\rho_S$ is the solar array specific mass per surface unit area and the constant $k_S$ represents the margin on solar array mass.

The same applies to the mirror's mass:

$$m_M = k_M \rho_M \left( A_d + A_{M_1} + 2 A_{M_2} \right) \qquad (11)$$

where $\rho_M$ is the mirror specific mass per unit area, $k_M$ is the margin on mirror mass, $A_{M1}$ is the area of the primary mirror and $A_{M2}$ and $A_d$ are the areas of the secondary and directional mirror respectively. They are defined as:

$$A_{M_2} = 0.01 A_{M_1}$$
$$A_d = \frac{A_{M_1}}{C_r} \qquad (12)$$

where $C_r$ is the concentration ratio, i.e. the ratio between the solar power density on the solar concentrator and that of the spot area on the asteroid. The mass of the laser is proportional to its output power:

$$m_L = k_L \rho_L P_L \eta_L \qquad (13)$$

where $\rho_L$ is the laser specific mass per input unit power, $k_L$ is the margin on laser mass and the input power $P_L$ depends on the solar input $P_{in}$ and the efficiency of the solar array $\eta_{SA}$:

$$P_L = \eta_{SA} P_{in} A_{M_1} \qquad (14)$$

Finally the total mass of the spacecraft is computed by adding a fixed mass fraction for the propellant:

$$m_{sc} = m_{dry} + 1.1 m_p = m_{dry} + 1.1 MF_p m_{dry} \qquad (15)$$

where $MF_p$ is the mass fraction for propellant and the factor 1.1 accounts for the mass of the tanks. The total mass of the formation is simply:

$$m_{sys} = n_{sc} m_{sc} \qquad (16)$$

and the global conversion efficiency of the laser system is given by:

$$\eta_{sys} = \eta_L \eta_{SA} \eta_P \varepsilon_M \qquad (17)$$

where $\eta_L$, $\eta_{SA}$, $\eta_P$, are the efficiency of the Laser, solar arrays and power bus respectively and $\varepsilon_M$ is the emissivity of the mirror. The constants $k_{dry}$, $k_S$, $k_M$, $k_L$ represent system margins that are chosen according to standard practice in space systems engineering and to design maturity (Wertz and Larson 1999). For example, for the dry mass a 20% margin (i.e. $k_{dry}$=1.2) is used since this is what is normally done in a preliminary mission design study; for the solar arrays, a 15% margin is deemed adequate given the maturity reached by the related technology; for the mirror mass instead, a higher value of 25% was preferred; finally, given the fact that high power lasers for space applications are still in their infancy, a 50% margin must be used for the laser (see Table 1). Margins are used when uncertainties are not quantified exactly. In the following, therefore, margin parameters will be equal to 1 when uncertainties are quantified through Evidence Theory.

| $k_{dry}$ | $k_S$ | $k_M$ | $k_L$ |
|---|---|---|---|
| *1.2* | 1.15 | 1.25 | 1.5 |

**Table 1** System design margins

One of the critical aspects of the design of the laser ablation system is that the quantities $\eta_L$, $\eta_{SA}$, $\rho_R$, $\rho_L$ and $\rho_M$ are poorly known. This is due to the fact that some of the related technologies are still in an early development stage. In particular, the efficiency and mass of the laser for space application are considered to be quite uncertain. As a matter of fact, there are two methods for powering the laser: in *direct pumping*, the solar energy is used to directly excite the electrons thereby generating the laser beam; on the other hand, in *indirect pumping*, the energy is first converted into electrical power, which then powers a semiconductor laser. Currently, high efficiency (up to 35%) directly pumped lasers have been discussed at a theoretical level while existing systems achieve only a few percent of power efficiency (Vasile et al. 2009b). Indirect pumping, instead, has shown very good performance albeit mainly in non-space applications and with lower power outputs. For indirect pumping systems, there is quite some uncertainty on the energy conversion efficiencies that will be achieved in the short or medium term. Efficiencies around 40-50% should be easily attainable even with current proven technology (combining semiconductor laser with fibres) but some laboratory tests have suggested that much higher values, around 65%, are probably achievable, assuming over 80% wall-plug efficiency of the semiconductors and over 80% of the fibres (Vasile et al. 2009b). Solar arrays are also a critical factor in the performance of an indirect pumped laser system. Recent advances in multiple junction cell technology have allowed for efficiencies close to 30% but it is not totally unrealistic to expect that near future improvements will move this threshold as high as 40-50% under concentrated light with partial efficiency recovery through thermocouples.

A third critical element is the radiator. As a matter of fact, given the relatively low power conversion efficiency of the solar arrays-laser combination (from ~10% to ~30% at best), most of the input solar power is rejected as heat and therefore must be dissipated by the radiators. While well proven, high emissivity, radiator technology is already available, the problem lays in the weight per emitting area for large systems. While for small radiator this is around 1 $kg/m^2$, for large surfaces this could be as high as 4 $kg/m^2$ (Vasile et al. 2009b). It is clear that these wide ranges on many different parameters can considerably affect the overall size of the laser system and consequently the mass of the laser formation to be put in orbit. At the same time, the lack of detailed knowledge on the physical characteristics of the NEO can markedly affect the system's capability in sublimating enough surface material as to generate enough thrust to deviate the asteroid.

The performance index which is output by the system model is the total system mass of the Laser satellite formation $m_{sys}$. The input design parameters are the number of spacecraft $n_{sc}$, the diameter of the primary mirror $d_{M1}$ and the concentration ratio $C_r$. As already mentioned the parameter subjected to uncertainties are $\eta_L$, $\eta_{SA}$, $\rho_R$, $\rho_L$ and $\rho_M$.

## 4. Deflection Action Model

As shown in previous works (Sanchez, Colombo et al. 2009; Vasile et al. 2009a; Maddock and Vasile 2008), the yield of the ablation process can be modelled with the simple energy balance (assuming no ionisation):

$$\frac{dm_{exp}}{dt} = 2n_{sc}v_{rot} \int_{y_0}^{y_{rot}} \int_{t_{in}}^{t_{out}} \frac{1}{E_{sub}} (P_{in} - Q_{rad} - Q_{cond}) dt dy \qquad (18)$$

where, $dm_{exp}/dt$ is the mass flow rate of sublimated material, $n_{sc}$ is the number of spacecraft in the formation, $v_{rot}$ is the linear velocity of the asteroid surface due to its rotation, $E_{sub}$ is the enthalpy of sublimation. The input power per unit area from the laser is:

$$P_{in} = \eta_{sys} C_r (1 - \varsigma_A) S_0 \left(\frac{r_{AU}}{r_A}\right)^2 \qquad (19)$$

where $\varsigma_A$ is the albedo of the asteroid, $S_0 = 1367 \, W/m^2$ is the solar flux at *1 AU*, $r_{AU}$ is the astronomical unit and $r_A$ is the Sun-asteroid distance. Here the assumption is that the amount of reflected laser light is comparable to the amount of reflected visible light. For a highly effective volumetric absorber, experimental evidence has shown that for given asteroid analogue target materials – sandstone, olivine, and a porous composite mixture – that the majority of the incoming laser intensity is absorbed rather than reflected. Energy is emitted in the form of an extended, but contained, exhaust of gas and ejecta. This has been demonstrated for a *90 W* continuous wave laser operating at a frequency of *808 nm* (Gibbings, Vasile et al. 2011a, 2011b). The heat loss due to black body radiation is:

$$Q_{rad} = \sigma \varepsilon_{bb} T^4 \tag{20}$$

where $\sigma$ is the Stefan-Boltzmann constant, $\varepsilon_{bb}$ is the black body emissivity, $T$ is the asteroid surface temperature. The loss due to thermal conduction is expressed as (Sanchez, Colombo et al. 2009):

$$Q_{cond} = (T_{subl} - T_0) \sqrt{\frac{c_A k_A \rho_A}{\pi t}} \tag{21}$$

with $T_{sub}$ as the temperature of sublimation of the surface material and $c_A$, $k_A$ and $\rho_A$ as its specific heat, thermal conductivity and density respectively. The ablation-induced acceleration can therefore be calculated as:

$$\mathbf{f}_{sub} = \frac{\Lambda \bar{v} \dot{m}_{exp}}{m_A} \hat{\mathbf{v}}_A \tag{22}$$

where $\hat{\mathbf{v}}_A$ is the unit vector along the NEO heliocentric velocity, $\Lambda \approx \frac{2}{\pi}$ is the scattering factor that assumes that the plume is uniformly distributed over an angle of *180 deg*, $m_A$ is the asteroid mass and $\bar{v}$ is the average velocity of the ejecta:

$$\bar{v} = \sqrt{\frac{8 k_B T_{subl}}{\pi M_{Mg2SiO_4}}} \tag{23}$$

where $k_B$ is the Boltzmann constant and $M_{Mg2SiO_4}$ is the molecular mass of Forsterite. Note that no ionization model is considered here. This assumption is consistent with the sublimation model in Kahle et al. where the power density is analogous to the one in this paper. A more accurate model is out of the scope of this paper, on the other hand the paper proposes a methodology to model and propagate uncertainties in order to evaluate the impact on the quantities of interest, such as the achievable miss distance. An unmodelled component has to be regarded as a source of model uncertainty. More specifically, the incident laser energy absorption and the expansion of the gas depend on the level of ionization (see Phipps et al. 2010). An uncertainty on energy absorption and gas expansion is equivalent to adding uncertainty to the sublimation Enthalpy and to the parameters defining the expansion velocity, as it will be presented in the next section.

The thrust model needs to be completed with a suitable model of the contamination of the optics. In fact the plume of gas and debris coming from the ablation process is expected to flow and impact the spacecraft. The contamination model used in this paper is the one developed by Kahle et al. (2006) and further elaborated in (Vasile and Maddock 2010). This model assumes that the sublimation of asteroid's surface is analogous to the generation of tails in comets and that the plume will expand as the exhaust gases of a rocket engine (as shown in Fig. 4). Note that, such a model is not strictly consistent with the hemispherical scattering model used for computing the ablation thrust. Moreover, experimental data (Gibbings, Vasile et al. 2011b) is showing that neither the hemispherical model nor the one by Kahle et al., shown in Fig. 4, accurately represent the expansion of the plume. However, they are used in the present work because each represents the worst case condition for thrust generation and mirror contamination respectively. The density of the expelled gas plume is computed as:

$$\rho_{exp} = j_C \frac{\dot{m}_{exp}}{\bar{v} A_{spot}} \left( \frac{d_{spot}}{2 r_{S/SC} + d_{spot}} \right)^2 (\cos \Theta)^{\frac{2}{\kappa - 1}} \tag{24}$$

where $j_C = 0.345$ is the jet constant, $\kappa = 1.4$ is the adiabatic index, $A_{spot}$ and $d_{spot}$ are respectively the area and diameter of the Laser spot on the asteroid; $r_{S/SC}$ is the norm of the distance vector of the spacecraft with respect to the spot on the asteroid. $\Theta$ is given by:

$$\Theta = \frac{\pi \varphi}{2 \varphi_{max}} \tag{25}$$

In the Hill reference frame $\mathbf{r}_{S/SC}$ is defined as:

$$r_{S/SC} = \begin{bmatrix} x - r_{ell} \sin \theta_{v_A} \\ y - r_{ell} \cos \theta_{v_A} \\ z \end{bmatrix} \tag{26}$$

where the radius of the ellipsoid is:

$$r_{ell} = \frac{a_I b_I}{\sqrt{\left(b_I \cos\left(\omega_A t + \theta_{v_A}\right)\right)^2 + \left(a_I \sin\left(\omega_A t + \theta_{v_A}\right)\right)^2}} \qquad (27)$$

*x,y* and *z* are the coordinates of the spacecraft with respect to the asteroid in the Hill reference frame, as shown in Fig. 6,and $a_I$ and $b_I$ are the axes of the ellipsoid (the asteroid is assumed to be a rotation ellipsoid),.

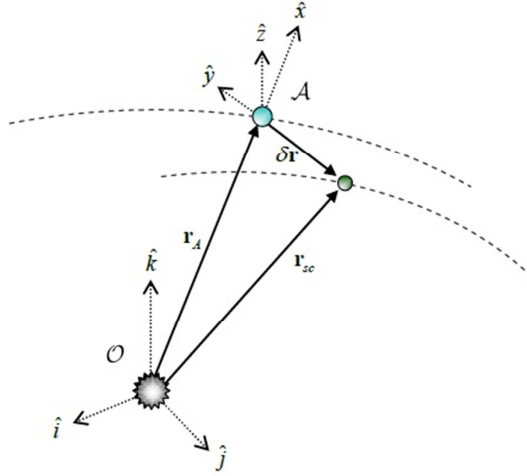

**Fig. 6** Hill reference frame

The asteroid is assumed to be spinning around the z-axis with angular velocity $\omega_A$. $\theta_A$ is the elevation of the spot over the y-axis. The model also assumes that all the particles impacting the mirror condense and stick to it. The variation of the thickness of the contamination layer on the mirror is thus computed as:

$$\frac{dh_{cond}}{dt} = \frac{2\bar{v}\rho_{\exp}}{\rho_{layer}} \cos\psi_{vf} \qquad (28)$$

where the layer density $\rho_{layer}$ is 1 $g/cm^3$. The speed of the ejecta is multiplied by 2 to account for the gas expansion in a vacuum. $\psi_{vf}$ is the view factor taken as the angle between the normal of the mirror and the incident flow of gas. Finally, the power irradiated on the asteroid's surface is multiplied by a degradation factor $\tau$:

$$\tau = \exp\left(-2\eta h_{cond}\right) \qquad (29)$$

where $\eta = 10^4$ $cm^{-1}$ is the absorption coefficient for Forsterite.

It is important to observe that, according to the relative motion as in Fig. 4, the mirrors would be exposed to the plume only for roughly half the period of the orbit of the asteroid, i.e. when the spacecraft has positive x coordinate.

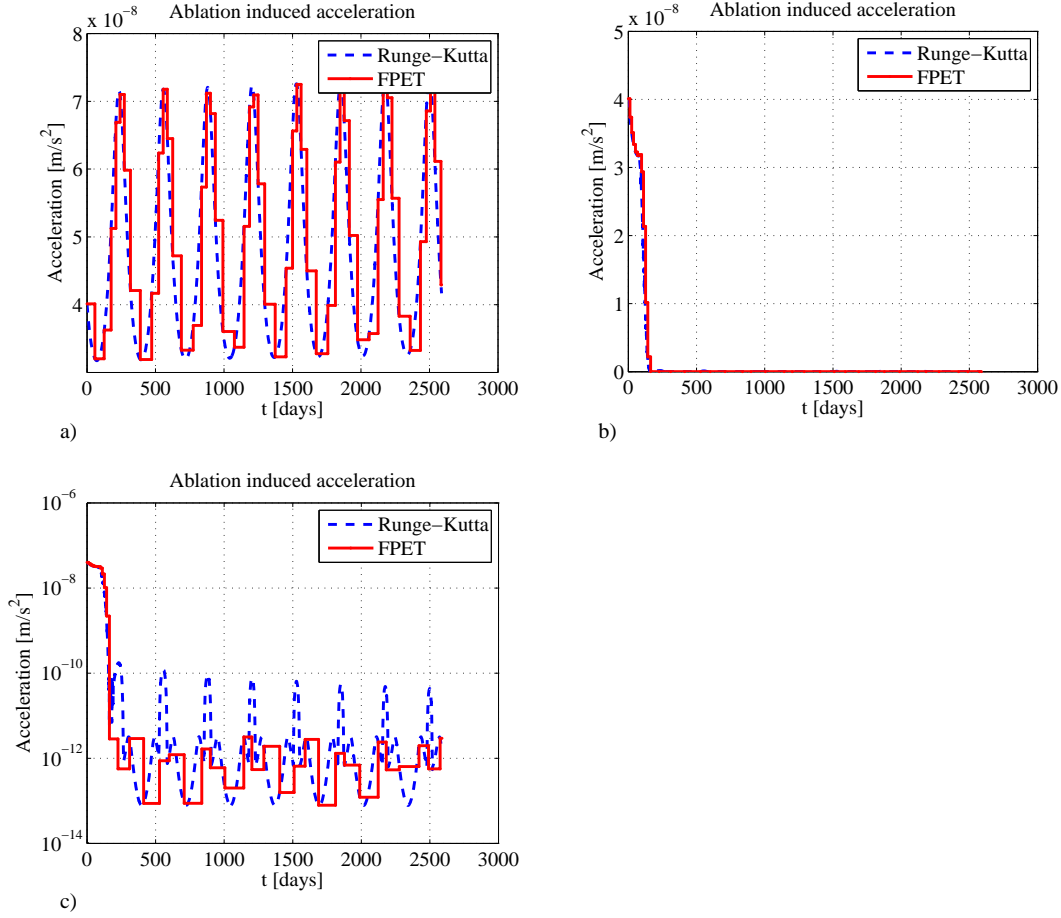

**Fig. 7** Typical acceleration profile: a) without contamination b) with contamination c) with contamination (semi-logarithmic scale)

Fig. 7a shows a typical acceleration profile computed without considering the contamination of the mirror. The figure compares the profile obtained from numerical integration of the trajectory and ablation models with a high order Runge-Kutta method, with the one obtained with analytical propagation with FPET. The periodic behaviour is due to NEO's motion around the Sun which accounts for oscillations in the solar flux captured by the primary mirror. The two integrations are in good agreement and the difference is due to Eq. (6). Fig. 7b and Fig. 7c show the same case but with the introduction of the contamination model in (Vasile and Maddock 2010). One can see that the amplitude of the acceleration oscillation decreases by more than two orders of magnitude already during the first revolution around the Sun and then stabilises at around $10^{-11}$-$10^{-13}$ $m/s^2$ for the rest of the trajectory.

From Fig. 7 it is important to observe that the FPET propagation approximates very accurately the acceleration profile when its magnitude is high during the first revolution and less correctly when it is decayed for the remainder of the trajectory. This will not affect the accuracy on the computation of the impact parameter since the contribution of the first part will be much more relevant than the second, which will be almost negligible.

As will be detailed in Section 5.1, from an analysis of the literature on NEO, one can observe a considerable variability of the physical parameters of asteroids, in particular $E_{sub}$, $T_{sub}$, $c_A$, $k_A$ and $\rho_A$, which are at the same time quite controversial and very critical to the laser ablation system design.

All these sources of uncertainty are of epistemic nature as they correspond to the present lack of knowledge on the asteroid physical properties. Due to the nature of the uncertainty, probability theory would be inadequate to model and quantify its value, therefore it is here proposed to use Evidence Theory to build a correct uncertainty model and introduce it in the combined optimal design of the deflection and spacecraft system.

## 5. Uncertainty Quantification

Evidence Theory, or Dempster-Shafer Theory, is a mathematical framework to model epistemic uncertainty and can be interpreted as a generalisation of classical probability theory (Klir and Smith 2001). In this sense, Evidence theory, is able to model both aleatory (i.e. related to stochastic processes) and epistemic (i.e. due to lack of knowledge) uncertainties. Differently from probability theory, where a probability distribution is used, in Dempster-Shafer theory an uncertain parameter $u_1$ can be modelled with one or more uncertain intervals $U^i_1$, each with its associated confidence level, also defined as Basic Probability Assignment (BPA):

$$U^i_1 = \left\{ \forall u_1 : u_1 \in [\underline{u}_{1_i}, \overline{u}_{1_i}] \right\}; \quad BPA(U^i_1) \in [0,1] \qquad (30)$$

Moreover, while in a standard probability distribution the integral over its domain of existence should be equal to one, the equivalent condition in Evidence Theory is less strict:

$$\sum_i BPA(U^i_1) + \sum_{i,j} BPA(U^i_1 \cup U^j_1) = 1 \qquad (31)$$

which translates into the fact that the intervals can not only be disconnected, but also overlapping. When dealing with multiple uncertain parameters, the uncertain space is defined by the Cartesian product of the single mono-dimensional uncertain intervals. A single multidimensional box, whose edges are the uncertain intervals for each uncertain parameter, is called *focal element* and its BPA is computed as:

$$BPA\left((u_1, u_2) \in [\underline{u}_{1_i}, \overline{u}_{1_i}] \times [\underline{u}_{2_j}, \overline{u}_{2_j}]\right) = BPA\left(u_1 \in [\underline{u}_{1_i}, \overline{u}_{1_i}]\right) \cdot BPA\left(u_2 \in [\underline{u}_{2_j}, \overline{u}_{2_j}]\right) \qquad (32)$$

Evidence theory also uses two complementary quantities to measure the cumulative confidence, or belief, in a given proposition: *Belief* and *Plausibility*. To explain their meaning, let us consider a performance parameter $y$ which is a function $f$ of the design parameters **x** and of the uncertain parameters **u**. The set of all $y$ which are below a certain threshold $\nu$ is defined as:

$$Y_\nu = \{ y : y = f(\mathbf{x}, \mathbf{u}) < \nu, \mathbf{x} \in D, \mathbf{u} \in U \} \qquad (33)$$

then the Belief and Plausibility associated to the proposition $y < \nu$ are:

$$\begin{aligned} Bel(Y_\nu) &= \sum_{j \in I_B} BPA(U^j) \\ Pl(Y_\nu) &= \sum_{j \in I_P} BPA(U^j) \end{aligned} \qquad (34)$$

with

$$\begin{aligned} I_B &= \left\{ j : U^j \subset f^{-1}(Y_\nu) \right\} \\ I_P &= \left\{ j : U^j \cap f^{-1}(Y_\nu) \neq 0 \right\} \end{aligned} \qquad (35)$$

It should be noted that $I_B$ is always a subset of $I_P$, i.e. $I_B \subseteq I_P$ and in this sense Belief and Plausibility can be interpreted as respectively the lower and upper boundary for the likelihood of an event. Differently from the probability of an event and its opposite, Belief and Plausibility are not strictly complementary. Instead the following relationships apply:

$$\begin{aligned} Bel(A) + Bel(\overline{A}) &\leq 1 \\ Pl(A) + Pl(\overline{A}) &\geq 1 \\ Bel(A) + Pl(\overline{A}) &= 1 \end{aligned} \qquad (36)$$

The next section will describe the procedure for defining the uncertain intervals and the focal elements.

## 5.1 Construction of the Uncertain Intervals

In this section, the uncertain intervals and the associated BPA for each uncertain parameter are defined. Moreover, it will be simulated the situation in which the estimates about the uncertain intervals and their associated confidence come from different sources. In order to do this, in this study the assumption is that the values of uncertain physical and technological parameters stem from the opinion of three different experts, as reported in Table 2, Table 3 and Table 4. Each expert expresses its own opinion on the uncertain intervals and assigns a personal confidence level to each of them. The confidence level represents the perception that experts have in their own level of knowledge. The opinions of the three experts could also be in disagreement with each other. This disagreement can be manifold. In the first instance, the experts can have different opinions on the amplitude of the interval itself and therefore propose slightly different boundaries. Secondly, even if the intervals proposed by different experts are the same, they can associate to them a different confidence and therefore estimate different BPAs. Moreover, some experts can also give a very generic indication that the given parameter can oscillate between a minimum and maximum value with equal confidence, which corresponds to giving a single wide interval with BPA equal to 1. And last, the expert can have no opinion at all on some quantities.

For the technological parameters $\eta_L$, $\eta_{SA}$, $\rho_R$, $\rho_L$ and $\rho_M$, the three experts behaves as follows. Regarding the laser efficiency, *expert a* in Table 2 is rather conservative and assigns a high confidence of 70% to the proposition that the efficiency will be between 40% and 50%; he/she is less confident about the possibility of achieving efficiencies comprised between 50% and 60% and therefore the related probability assignment is 30%. *Expert b*, in Table 3, on the other hand is probably more realistic and assigns only 30% confidence to the interval of 40-50% efficiency, while giving 60% to the 50-60% efficiency interval and finally introducing another interval between 60% and 66.4% with a confidence of 10%. *Expert c,* in Table 4, is very optimistic about future developments of lasers and therefore assigns 100% confidence to the statement that lasers could reach efficiencies between 55% and 66.4%. For the laser specific mass, *expert a* gives 40% confidence about the specific mass being comprised between 0.005 and 0.01 *kg/W* while is more oriented towards higher specific masses in the interval of 0.01-0.02 *kg/W* and therefore assigns 60% confidence to the latter. *Expert b*, on the other hand, is convinced that lightweight laser systems are possible and therefore assign 100% to the 0.01-0.02 *kg/W*. *Expert c* does not give any opinion on this topic (reported as *n/a* in the table). For the solar array efficiency, *expert a* is again rather sceptical and proposes only one interval between 20% and 30%, obviously with 100% confidence. *Expert b* suggests only a 40% confidence for the 20-30% efficiency range and instead assigns a 60% confidence about achieving higher efficiencies comprised between 30% and 50%. *Expert c* again doesn't express any opinion on the topic (reported as *n/a* in the table). Regarding the mirror specific mass, *expert a* is equally oriented towards values between 0.1 and 0.3 $kg/m^2$ and 0.3 and 0.5 $kg/m^2$, therefore confidence will be 50% for both. *Expert b* again proposes only one interval with 100% confidence for values ranging from 0.3 and 0.5 $kg/m^2$. *Expert c* instead is very optimistic about the development of lightweight mirrors with specific masses between 0.01 and 0.05 $kg/m^2$. Finally for the radiator, *expert a* suspects that radiator specific mass will be higher for large radiators like those envisioned for laser ablation spacecraft and therefore suggests 40% for values comprised in the 1-2 $kg/m^2$ and 60% for values between 2 and 4 $kg/m^2$. *Expert b* doesn't give an opinion on the topic (reported as *n/a* in the table) while *expert c* gives a generic indication that the mirror specific mass will surely be between 1 and 3 $kg/m^2$.

As already pointed out in Section 4, physical properties can differ considerably from one asteroid to the other. At the same time, different sources report different physical parameters for the same asteroid. Moreover, data is currently limited to ground based observations and a limited number of fly-by missions to only a few NEOs, such as Eros, Itokawa, Steins and Lutetia. However, these missions demonstrated that the fundamental nature, composition and geometries of NEOs are highly variable. Any generic group of physical characteristics can introduce a significant error within the analysis. Furthermore substantial error bars in $T_{sub}$, $c_A$, $k_A$ and $\rho_A$ also exist from the inferred spectra analysis and shape regularity – including period of rotation, form and shape model, and surface properties (Britt et al. 2002; Pieters and McFadden, 1994; Price 2004). For example, available source show a range of two orders of magnitude for the sublimation enthalpy: it is as low as $2.7 \cdot 10^5$ *J/kg* for some rare E type asteroids composed by carbonaceous and Enstatile Chondrites while it can reach $1.9686 \cdot 10^7$ *J/kg* for some S type asteroids with Olivine composition. For Silicum based bodies, the values are intermediate, around $5 \cdot 10^6$ *J/kg*. In this respect, for example, *expert a* gives 100% confidence to enthalpy being generically comprised between values as low as $2.7 \cdot 10^5$ and as high as $6 \cdot 10^6$ *J/kg*. *Expert b* gives more details, proposing only 20% confidence for a lower range between $2.7 \cdot 10^5$ and $10^6$ *J/kg* for Chondritic objects and assigning instead 80% confidence to enthalpies comprised in the $10^7$-$1.9686 \cdot 10^7$ *J/kg* typical of S-type Olivine asteroids. *Expert c*, while agreeing on the boundaries of this interval, assigns only 30% confidence to it and also is more persuaded about a different lower interval between $4 \cdot 10^6$ and $6 \cdot 10^6$ *J/kg*, to which he assigns a 70% confidence. Analogously, for the specific heat, most sources reported values between 500 and 600 *J/(kg·K)*, which are typical of Olivine-based S type asteroid but also of M and C types such as Lutetia and Mathilde. It is interesting to note, however, that in some cases like the E type asteroid Steins the estimates can range from 470 up to over 750 *J/(kg·K)*. Thus, *expert a* suggests two uncertain intervals: the first from 375 to 470 *J/(kg·K)* with 30% confidence, and the second one from 470 to 600 *J/(kg·K)* with 70% confidence. Also *expert b* proposes this latter range, but with 40% confidence only. He also proposes a higher interval from 600 up to 750 *J/(kg·K)* with 60% confidence. *Expert c* gives a generic indication that the specific heat will be between 470 and 750 *J/(kg·K)*. For the thermal conductivity, the range spans two orders of magnitude: for common S-type, Olivine bodies and for some E type asteroids it is around 1.47-1.6 *W/(m·K)*; it is as low as 0.2 *W/(m·K)* for others like M-type Lutetia and C-type Mathilde. In this sense, *expert a* assigns 20% confidence to an interval to a low interval for relatively rare M/C-type bodies with conductivities comprised between 0.2 and 0.5 *W/(m·K)*. On the other hand, he/she gives 80% to the assumption that the conductivity will be between 1.47 and 1.6 *W/(m·K)*. *Expert b* is again rather generic giving just a minimum of 0.2 *W/(m·K)* and maximum of 2 *W/(m·K)*. *Expert c* is unable to give an opinion (reported as *n/a* in the table). Regarding the density, sources report values comprised between 1100 and 2000 *kg/m³* for most C-type asteroids, and between 2000 and 3700 *kg/m³* for S-types and some M-type ones. According to this, *expert a* thinks that S-type objects will be more common and therefore assigns 70% to the latter interval and 30% to the former. This time too, *expert b* is very vague, giving indications of a lower bound at 1100 *kg/m³* and an upper at 3700 *kg/m³*. *Expert c* disagrees with the lower limit and sets it at 2000 *kg/m³* instead. Finally, the sublimation temperature shows a more limited variability, with values around 1700 *K* for S-type and up to 1812 *K* for other examples. This small variability is also reflected in the experts' opinion, since *expert a* assumes the values related to S-type asteroids, between 1700 *K* and 1720 *K*, with 100% confidence. *Expert b* proposes a range spanning 1720-1812 *K*, again with 100% confidence, while *expert c* proposes a wider range from 1700 *K* to 1812 *K*.

The three sources of information are data-fused following a similar procedure to the one described by Oberkampf and Helton (2002). As a representative example, the procedure is here applied to the data-fusion of the estimates concerning the laser efficiency. As already discussed, the opinions given by three experts are:
  a. Conservative opinion: "The Laser efficiency will be between 40% and 50% with 70% confidence and between 50% and 60% with 30% confidence".
  b. Realistic opinion: "The Laser efficiency will be between 40% and 50% with 30% confidence, between 50% and 60% with 60% confidence and between 60% and 66.4% with 10% confidence".
  c. Optimistic opinion: "The Laser efficiency will be between 55% and 66.4% with 100% confidence".

These statements, in mathematical terms can be written as:

a. $\begin{cases} {}^aU^1 = [0.4, 0.5] & BPA({}^aU^1) = 0.7 \\ {}^aU^2 = [0.5, 0.6] & BPA({}^aU^2) = 0.3 \end{cases}$

b. $\begin{cases} {}^bU^1 = [0.4, 0.5] & BPA({}^bU^1) = 0.3 \\ {}^bU^2 = [0.5, 0.6] & BPA({}^bU^2) = 0.6 \\ {}^bU^3 = [0.6, 0.664] & BPA({}^bU^3) = 0.1 \end{cases}$

c. ${}^cU = [0.55, 0.664] \quad BPA({}^cU) = 1$

Then, to represent and then combine the data given by the three experts, for each of them a matrix is constructed as follows (Oberkampf and Helton 2002):
  1. First, one has to list all the possible values the experts propose as lower and upper boundaries for the uncertain intervals. In this case the lower boundaries are $[0.4 \quad 0.5 \quad 0.55 \quad 0.6]$ and upper boundaries are $[0.5 \quad 0.6 \quad 0.664]$.
  2. Then, a lower triangular matrix $A_i$ is defined for each source of information, which has as many columns as the possible lower boundaries and as many rows as the possible upper boundaries. Thus, each element of this lower triangular matrix represents a certain interval with its lower and upper limits. If the expert has associated a confidence level to that interval, then the element of the matrix assumes that value and is zero otherwise. For example, the matrix for *expert a* will have the following structure:

|       | 0.4 | 0.5 | 0.55 | 0.6 |
|-------|-----|-----|------|-----|
| 0.5   | 0.7 | 0   | 0    | 0   |
| 0.6   | 0   | 0.3 | 0    | 0   |
| 0.664 | 0   | 0   | 0    | 0   |

In the present case, the three matrices are as follows.

a. $A_a = \begin{bmatrix} 0.7 & 0 & 0 & 0 \\ 0 & 0.3 & 0 & 0 \\ 0 & 0 & 0 & 0 \end{bmatrix}$

b. $A_b = \begin{bmatrix} 0.3 & 0 & 0 & 0 \\ 0 & 0.6 & 0 & 0 \\ 0 & 0 & 0 & 0.1 \end{bmatrix}$

c. $A_c = \begin{bmatrix} 0 & 0 & 0 & 0 \\ 0 & 0 & 0 & 0 \\ 0 & 0 & 1 & 0 \end{bmatrix}$

At this point the three sets of intervals can be combined into a single one by computing the weighted average of matrices as:

$$A = \frac{k_a A_a + k_b A_b + k_c A_c}{3} \quad (37)$$

where $k_a$, $k_b$, and $k_c$ are weights which can be defined arbitrarily in order to give different influence to each source of information. In this case, all sources are given the same importance and therefore the weights are all set to 1. The resulting matrix is therefore:

$$A = \begin{bmatrix} 0.3333 & 0 & 0 & 0 \\ 0 & 0.3 & 0 & 0 \\ 0 & 0 & 0.3333 & 0.0333 \end{bmatrix}$$

from which one derives the uncertain intervals as:

$$U^1 = [0.4, 0.5] \quad BPA(U^1) = 0.3333$$
$$U^2 = [0.5, 0.6] \quad BPA(U^2) = 0.3$$
$$U^3 = [0.55, 0.664] \quad BPA(U^3) = 0.3333$$
$$U^4 = [0.6, 0.664] \quad BPA(U^4) = 0.0333$$

A similar procedure was followed for the remaining nine uncertain parameters, leading to the results reported in Table 5 and Table 6. Note that information fusion of different sources for this specific case is still an open problem. The use of a weighted average is only one possibility. A thorough analysis of the right information fusion technique is out the scope of this paper and will be addressed in future works.

|                   | Lower    | Upper    | BPA |                     | Lower | Upper | BPA |
|-------------------|----------|----------|-----|---------------------|-------|-------|-----|
| $c_A$ [J/KgK]     | 375      | 470      | 0.3 | $\eta_L$            | 0.4   | 0.5   | 0.7 |
|                   | 470      | 600      | 0.7 |                     | 0.5   | 0.6   | 0.3 |
| $k_A$ [W/mK]      | 0.2      | 0.5      | 0.2 | $\eta_{SA}$         | 0.2   | 0.3   | 1   |
|                   | 1.47     | 1.6      | 0.8 |                     |       |       |     |
| $\rho_A$ [kg/m³]  | 1100     | 2000     | 0.3 | $\rho_M$ [kg/m²]    | 0.1   | 0.3   | 0.5 |
|                   | 2000     | 3700     | 0.7 |                     | 0.3   | 0.5   | 0.5 |
| $T_{sub}$ [K]     | 1700     | 1720     | 1   | $\rho_L$ [kg/W]     | 0.005 | 0.01  | 0.4 |
|                   |          |          |     |                     | 0.01  | 0.02  | 0.6 |
| $E_{sub}$ [J/kg]  | 2.7·10⁵  | 6·10⁶    | 1   | $\rho$ [kg/m²]      | 1     | 2     | 0.4 |

|   |   |   | 2 | 4 | 0.6 |
|---|---|---|---|---|---|

**Table 2** Uncertain parameters estimates from *expert a*

|   | Lower | Upper | BPA |   | Lower | Upper | BPA |
|---|---|---|---|---|---|---|---|
| $c_A$ [J/KgK] | 470 | 600 | 0.4 | $\eta_L$ | 0.4 | 0.5 | 0.3 |
|   | 600 | 750 | 0.6 |   | 0.5 | 0.6 | 0.6 |
|   |   |   |   |   | 0.6 | 0.664 | 0.1 |
| $k_A$ [W/mK] | 0.2 | 2 | 1 | $\eta_{SA}$ | 0.2 | 0.3 | 0.4 |
|   |   |   |   |   | 0.3 | 0.5 | 0.6 |
| $\rho_A$ [kg/m$^3$] | 1100 | 3700 | 1 | $\rho_M$ [kg/m$^2$] | 0.3 | 0.5 | 1 |
| $T_{sub}$ [K] | 1720 | 1812 | 1 | $\rho_L$ [kg/W] | 0.01 | 0.02 | 1 |
| $E_{sub}$ [J/kg] | $2.7 \cdot 10^5$ | $10^6$ | 0.2 | $\rho_R$ [kg/m$^2$] |   | n/a |   |
|   | $10^7$ | $1.9686 \cdot 10^7$ | 0.8 |   |   |   |   |

**Table 3** Uncertain parameters estimates from *Expert b*

|   | Lower | Upper | BPA |   | Lower | Upper | BPA |
|---|---|---|---|---|---|---|---|
| $c_A$ [J/KgK] | 470 | 750 | 1 | $\eta_L$ | 0.55 | 0.664 | 1 |
| $k_A$ [W/mK] |   | n/a |   | $\eta_{SA}$ |   | n/a |   |
| $\rho_A$ [kg/m$^3$] | 2000 | 3700 | 1 | $\rho_M$ [kg/m$^2$] | 0.01 | 0.05 | 1 |
| $T_{sub}$ [K] | 1700 | 1812 | 1 | $\rho_L$ [kg/W] |   | n/a |   |
| $E_{sub}$ [J/kg] | $4 \cdot 10^6$ | $6 \cdot 10^6$ | 0.7 | $\rho_R$ [kg/m$^2$] | 1 | 3 | 1 |
|   | $10^7$ | $1.9686 \cdot 10^7$ | 0.3 |   |   |   |   |

**Table 4** Uncertain parameters estimates from *Expert c*

|   | Lower | Upper | BPA |
|---|---|---|---|
| $c_A$ [J/KgK] | 375 | 470 | 0.1 |
|   | 470 | 600 | 0.3667 |
|   | 470 | 750 | 0.3333 |
|   | 600 | 750 | 0.2 |
| $k_A$ [W/mK] | 0.2 | 0.5 | 0.1 |
|   | 1.47 | 1.6 | 0.4 |
|   | 0.2 | 2 | 0.5 |
| $\rho_A$ [kg/m$^3$] | 1100 | 2000 | 0.1 |
|   | 2000 | 3700 | 0.5667 |
|   | 1100 | 3700 | 0.3333 |
| $T_{sub}$ [K] | 1700 | 1720 | 0.3333 |
|   | 1720 | 1812 | 0.3333 |
|   | 1700 | 1812 | 0.3333 |
| $E_{sub}$ [J/kg] | $2.7 \cdot 10^5$ | $10^6$ | 0.0667 |
|   | $2.7 \cdot 10^5$ | $6 \cdot 10^6$ | 0.3333 |
|   | $4 \cdot 10^6$ | $6 \cdot 10^6$ | 0.2333 |
|   | $10^7$ | $1.9686 \cdot 10^7$ | 0.3667 |

**Table 5** Uncertain intervals of NEO physical properties

|   | Lower | Upper | BPA |
|---|---|---|---|
| $\eta_L$ | 0.4 | 0.5 | 0.3333 |
| | 0.5 | 0.6 | 0.3 |
| | 0.55 | 0.664 | 0.3333 |
| | 0.6 | 0.664 | 0.0333 |
| $\eta_{SA}$ | 0.2 | 0.3 | 0.2 |
| | 0.3 | 0.5 | 0.3 |
| | 0.2 | 0.5 | 0.5 |
| $\rho_M$ [kg/m$^2$] | 0.3 | 0.5 | 0.5 |
| | 0.1 | 0.3 | 0.1667 |
| | 0.01 | 0.05 | 0.3333 |
| $\rho_L$ [kg/W] | 0.005 | 0.01 | 0.2 |
| | 0.01 | 0.02 | 0.8 |
| $\rho_R$ [kg/m$^2$] | 1 | 2 | 0.2 |
| | 1 | 3 | 0.5 |
| | 2 | 4 | 0.3 |

**Table 6** Uncertain intervals of technological parameters

## 6. Multi Objective Optimization Under Uncertainty

Once the uncertainties on system design and asteroid physical characteristics are defined, one can try to find the optimal design of the deflection system under uncertainty. The performance, i.e. the achieved deviation, needs to be maximised while minimising a measure of the cost of the mission, e.g. the mass into space. According to the spacecraft system model presented in previous sections, performance and cost can be optimised with respect to four design parameters: the diameter of the primary mirror $d_M$, the number of spacecraft $n_{sc}$, the warning time $t_{warn}$ (time from the beginning of the deflection action to the time of the expected impact with the Earth) and the concentration ratio $C_r$. The performance measure to be maximised is the impact parameter $b$, while the cost measure to be minimised is the total mass of the formation $m_{sys}$. This leads to a classical multiobjective optimisation problem. The impact parameter is computed by means of the deflection and ablation model detailed in Sec.2 and Sec.4 while the total system mass is derived as in Sec.3.

As a first step one can determine the set of Pareto optimal solutions for a fixed value of the uncertain parameters $\eta_L$, $\eta_{SA}$, $\rho_R$, $\rho_L$, $\rho_M$, $E_{sub}$, $T_{sub}$, $c_A$, $k_A$ and $\rho_A$. Their value was chosen according to the available literature (Britt et al. 2002; Pieters and McFadden, 1994; Price 2004) and are reported in Table 7. Moreover, since at this stage uncertainties are not yet accounted for with Evidence theory, system margins as in Table 1 are included in the model, in order to replicate the standard system engineering method to deal with uncertainty.

| NEO Physical properties | | Technological parameters | |
|---|---|---|---|
| Parameter | Value | Parameter | Value |
| $c_A$ [J/KgK] | 750 | $\eta_L$ | 0.6 |
| $k_A$ [W/mK] | 2 | $\eta_{SA}$ | 0.41 |
| $\rho_A$ [kg/m$^3$] | 2600 | $\rho_M$ [kg/m$^2$] | 0.1 |
| $T_{sub}$ [K] | 1800 | $\rho_L$ [kg/W] | 0.005 |
| $E_{sub}$ [J/kg] | 5·10$^6$ | $\rho_R$ [kg/m$^2$] | 1.4 |

**Table 7** Set of fixed values for uncertain parameters

The multi objective optimisation problem to be solved is:

$$\min_{\mathbf{x}\in D}\left[ m_{system}\left(\mathbf{x},\overline{\mathbf{u}}\right) \quad -b\left(\mathbf{x},\overline{\mathbf{u}}\right) \right] \quad (38)$$

where **x** is the design parameter vector comprising $\mathbf{x}=[d_M, n_{sc}, t_{warn}, C_r]^T$, for which the boundaries are in Table 8, and $\overline{\mathbf{u}}$ is the vector of uncertain parameters with values in Table 7. The impact parameter $b$ appears with the minus sign since it has to be maximised. For the solution of problem (38) system margins are introduced with the values in Table 1.

|            | Lower | Upper |
|------------|-------|-------|
| $d_M$ [m]  | 2     | 20    |
| $n_{sc}$   | 1     | 10    |
| $t_{warn}$ [yrs] | 1 | 8    |
| $C_r$      | 1000  | 3000  |

**Table 8** Boundaries for optimization parameters

Note that the presence of the discrete variable $n_{sc}$ makes this a mixed integer-nonlinear multiobjective optimisation problem. The optimisation problem is solved with MACS, a hybrid memetic stochastic algorithm (Vasile and Zuiani 2011).

When epistemic uncertainties are introduced through Evidence Theory the MOO problem (38) has to be reformulated in order to maximise the Belief in the optimal value of impact parameter and total system mass. Formally problem (38) would translate into the MOO under uncertainty:

$$\max_{\mathbf{x} \in D} Bel(-b(\mathbf{x},\mathbf{u}) < \nu_b)$$
$$\max_{\mathbf{x} \in D} Bel(m_{sys}(\mathbf{x},\mathbf{u}) < \nu_m)$$
$$\min \nu_b$$
$$\min \nu_m$$
(39)

The solution of problem would require the computation of the Belief value for different design parameters and for different values of the thresholds $\nu_b$ and $\nu_m$ for all possible values of the uncertain parameters $\mathbf{u}$ within the uncertain space $U$: for each $\mathbf{x}$, set (33) needs to be computed for each of the functions $b$ and $m_{sys}$ for different $\nu_b$ and $\nu_m$ respectively and the cumulative functions (34) need to be independently computed for both $b$ and $m_{sys.}$. The identification of the set (33) would need the computation of the max and min of $b$ and $m_{sys}$ over all the focal elements in $U$. However, the number of focal elements in $U$ is an exponential function of the number of uncertain parameters (Vasile and Croisard 2010) which translates into an exponentially increasing number of optimisation problems required to compute the cumulative quantities in (34). In practise, however, the full Belief and Plausibility curves are not required and one can study only the worst and best case scenarios.

The best case scenario corresponds to the design, uncertainty vectors and thresholds that yield a Plausibility equal to 0. Below this value of the thresholds the deflection mission is not possible assuming the available body of knowledge of spacecraft systems and asteroid physical properties. The worst case scenario corresponds to the design, uncertainty vectors and thresholds that yield a Belief equal to 1. Above this value of the thresholds the mission is certainly possible, given the current body of knowledge, but would be suboptimal.

The optimal design vector and thresholds that yield a Belief equal to 1 for all possible $\mathbf{u}$ in $U$ can be computed solving the following multiobjective *minmax* problem:

$$\min_{\mathbf{x} \in D} \left[ \max_{\mathbf{u} \in \bar{U}} m_{system}(\mathbf{x},\mathbf{u}) \quad \max_{\mathbf{u} \in \bar{U}} (-b(\mathbf{x},\mathbf{u})) \right]$$
(40)

In fact, for a given $\mathbf{x}$, the minimum possible threshold value corresponds to the maximum value of $m_{sys}$ and $-b$ over the whole uncertain space $U$, for which boundaries are reported in Table 5 and Table 6. Because the focal elements in $U$ can be overlapping or can be disconnected, the identification of the maximum of $m_{sys}$ and $-b$ might be problematic as one would need to explore each focal element independently and therefore face an exponential number of optimisation problems. In order to avoid this exponential complexity, all focal elements are collected, through an affine transformation, into the unit hypercube $\bar{U}$ such that they are not overlapping or disconnected.

The optimal design vector and thresholds that yield a Plausibility equal to zero for all possible $\mathbf{u}$ in $U$ can be computed by solving the following multiobjective *minmin* problem:

$$\min_{\mathbf{x} \in D} \left[ \min_{\mathbf{u} \in \bar{U}} m_{system}(\mathbf{x},\mathbf{u}) \quad \min_{\mathbf{u} \in \bar{U}} (-b(\mathbf{x},\mathbf{u})) \right]$$
(41)

Again as before the focal elements are mapped into the unit hypercube $\bar{U}$ and the search is run over $\bar{U}$. Note that, differently from the case of problem (38), system design margins are no longer needed and therefore the values for $k_{dry}$, $k_S$, $k_M$, $k_L$ are all set to 1.

In this paper, the two mixed integer optimisation problems (40) and (41) are solved with a variant of Multi-Agent Collaborative Search (MACS) (Vasile and Zuiani 2011). This variant is tailored specifically to the solution of multiobjective *minmin/minmax* problems. The standard MACS framework is used to explore the design space $D$ and solve the minimisation problem, i.e. generate new candidate design vectors $\mathbf{x_c}$ and select the ones that minimise the vector function:

$$\mathbf{J}(\mathbf{x}_c) = \left[ \max_{\mathbf{u} \in \bar{U}} m_{system}(\mathbf{x}_c, \mathbf{u}) \quad \max_{\mathbf{u} \in \bar{U}} (-b(\mathbf{x}_c, \mathbf{u})) \right]^T \quad (42)$$

The value of each component of the vector function $\mathbf{J}$ is the result of a single objective maximisation over the space of the uncertain parameters.

The maximisation subproblems in (42) are solved by running an evolutionary algorithm based on Inflationary Differential Evolution (IDEA) (Vasile et al. 2011a) for a fixed number of function evaluations. MACS was run for 30000 function evaluations with 10 individuals, of which 2 are explorers and perform a local search (see Vasile and Zuiani 2011 for details). The sub-cycles with IDEA where run for 250 function evaluations with 5 individuals. These settings were devised after a series of preliminary tests. The solution of problems (40) and (41) provides the intervals for both the performance and the design parameters. In particular, the worst case corresponds to the maximum Belief condition:

$$\bar{\mathbf{y}} = [\bar{\mathbf{x}}, \bar{\mathbf{u}}] = \arg \min_{\mathbf{x} \in D} \left[ \max_{\mathbf{u} \in \bar{U}} m_{system}(\mathbf{x}, \mathbf{u}) \quad \max_{\mathbf{u} \in \bar{U}} (-b(\mathbf{x}, \mathbf{u})) \right]$$
$$Bel(\bar{\mathbf{y}}) = 1 \quad (43)$$

The best case instead corresponds to the minimum Plausibility point:

$$\underline{\mathbf{y}} = [\underline{\mathbf{x}}, \underline{\mathbf{u}}] = \arg \min_{\mathbf{x} \in D} \left[ \min_{\mathbf{u} \in \bar{U}} m_{system}(\mathbf{x}, \mathbf{u}) \quad \min_{\mathbf{u} \in \bar{U}} (-b(\mathbf{x}, \mathbf{u})) \right]$$
$$Pl(\underline{\mathbf{y}}) = 0 \quad (44)$$

As a comparison, a *minmin* problem analogous to (44) is solved with the reintroduction of system design margins. Finally the 4 optimisation problems are considered both in the case with and without the contamination are solved. In summary, a total of 8 Pareto curves are generated, 4 each for the cases with and without the contamination:

1. *deterministic*, i.e. a bi-objective optimisation problem on $\mathbf{x} \in D$ as in (38). The system model does include the margins specified in Table 1 and constant values for uncertain parameters u are used as in Table 7. The problem is solved with the standard MACS.
2. *minmax*, bi-objective optimisation problem as in (40). The system model doesn't include margins. The problem is solved with the modified MACS.
3. *minmin*, bi-objective optimisation problem as in (41). The system model doesn't include margins. The problem is solved with the modified MACS.
4. *minmin with margins*, bi-objective optimisation problem as in (41). It's analogous to the previous one but this time the system model does include the margins specified in Table 1. The problem is solved with the modified MACS.

Fig. 8a and Fig. 8b report the Pareto fronts for the *deterministic*, *minmin* and *minmax* problems, with and without contamination respectively.

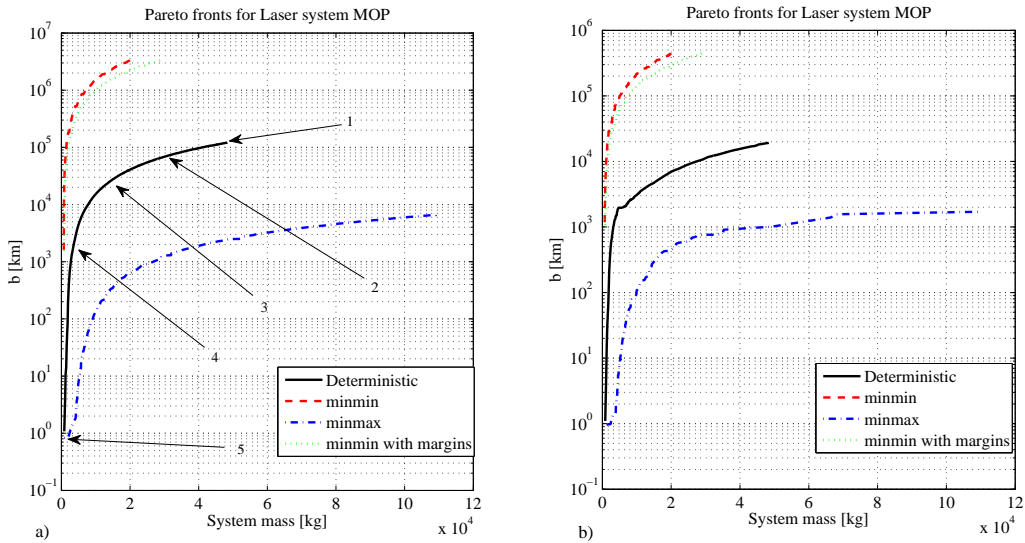

**Fig. 8** Multi objective optimization: Pareto fronts a) no contamination b) with contamination. *b* is represented in logarithmic scale

Qualitatively, the case with and without contamination are very similar but Fig. 8a shows that, without contamination, the best deviation achievable is one order of magnitude larger than in the case with contamination (see Fig. 8b). Since the system model is the same, the range of total system mass is the same in both cases.

The uncertainties in the input parameters translate into a difference between the *minmin* and *minmax* curves of about two orders of magnitude in attainable deviation and one order of magnitude in system mass. The achievable deviation easily reaches $10^5$-$10^6$ *km* in the best case scenario with a total formation mass below 30000 *kg*, while in the worst case scenario even with a system mass of $10^5$ *kg* the best achievable deviation does not exceed $10^4$ *km*. This issue is even more apparent in the case with mirror contamination in which the worst case deviation barely reaches $10^3$ *km*. It is important to point out that the huge variability in performance does not imply that the laser ablation is an unreliable deflection method as the type of uncertainty is epistemic. It implies instead that: given the present body of knowledge a reliable deflection mission would require a massive system in orbit, the potential margin for improvement would be considerable, current knowledge on this deflection method is too low to provide an exact quantification of its performance. Note also that the Pareto front for the case *minmin* with margins has higher system mass for the level of deviation attained with respect to the standard *minmin* case.

Fig. 9a and Fig. 9b show the distribution of optimal design solutions in the three case studies, without and with contamination respectively. The plots present the values only for three design variables, i.e. the diameter of the primary mirror, the number of spacecraft and the warning time. The concentration ratio is not reported because all the optimal design points show the maximum allowed concentration ratio allowed, i.e. 3000.

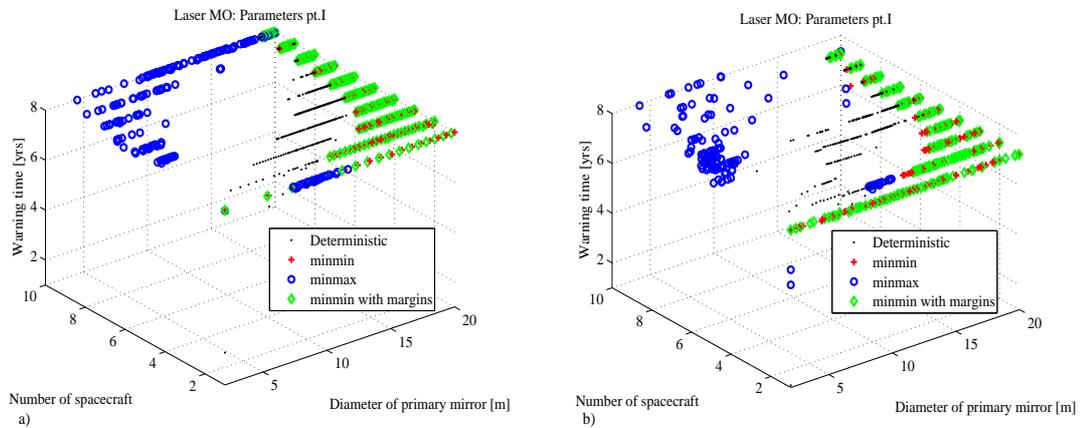

**Fig. 9** Multi objective optimization: Pareto sets a) no contamination b) with contamination

In Fig. 9 one can clearly identify two different families of design solutions in the *minmin* and *minmax* case. In the latter, solutions with a high number of spacecraft and a small diameter of the primary mirror are preferred. Arguably, many spacecraft are needed because the physical properties of the asteroid are such that inducing sublimation requires a large amount of power; at the same time the efficiency of the laser system is much lower and in particular the radiator mass per unit area is much higher and therefore it is convenient to have many smaller spacecraft, i.e. with a smaller primary mirror. Coherently with this, for diametrically opposite reasons, in the *minmin* case designs with few spacecraft with large concentrators are preferable. This result brings to an interesting general conclusion: for low performance components a monolithic system is suboptimal with respect to a disaggregated system as the mass of a monolithic system grows faster than the linear growth of the mass of the disaggregated counterpart. Note that, although redundancy was not modelled, the robust analysis suggests that a highly redundant system is preferable in the case of high uncertainty on the design parameters, as it would be logical to expect.

Finally one can note that in the case without contamination the maximum warning time of eight years is always optimal. This is easily explained given the fact that the magnitude of the thrust acceleration is relatively constant (albeit within a minimum and maximum values, see Fig. 7a) and therefore the longer this is acting on the NEO, the better. When the contamination of the mirrors is considered, then the optimal warning time is around 7.27 years. In this case, in fact, the acceleration profile essentially is reduced to a single large thrust impulse followed by a perturbation some orders of magnitude smaller than the initial peak (as shown in Fig. 7b and Fig. 7c). In this case, thus, the phasing of the initial pulse becomes extremely important (see Colombo et al. 2009).

### 6.1 Belief and Plausibility Analysis

To further analyse the influence of each individual uncertain parameter, five design points from the solution set of the deterministic case in Fig. 8a were selected. For each of them, the belief and plausibility curves for both the impact parameter $b$ and the system mass were reconstructed. The curves were computed with an algorithm based on the evolutionary binary tree technique in Vasile et al. (2011b):

1. Given the performance parameter $J_i$ and a constant design parameter vector $\bar{\mathbf{x}}$, the single objective optimisation problems:

$$\begin{aligned} \nu_{\min} &= \min_{\mathbf{u} \in \bar{U}} J_i(\bar{\mathbf{x}}, \mathbf{u}) \\ \nu_{\max} &= \max_{\mathbf{u} \in \bar{U}} J_i(\bar{\mathbf{x}}, \mathbf{u}) \end{aligned} \quad (45)$$

are solved with IDEA over the entire uncertain space given by the unit hypercube $\bar{U}$. This returns the upper and lower limit for the performance parameter.

2. $n_v$ values $v_j$ are defined equally spaced in the interval $\begin{bmatrix} \nu_{\min} & \nu_{\max} \end{bmatrix}$.

3. The initial unit hypercube $\bar{U}$ is partitioned in two sub-hypercubes $\bar{U}^1$ and $\bar{U}^2$. The "cut" is performed such that it coincides with the boundaries of adjacent focal elements which form the hypercube $\bar{U}$. Define $\Upsilon$ as the set of sub-hypercubes $\bar{U}^l$.

4. For each value of the threshold $v_j$, the following iterative procedure is performed:

   a. $\begin{aligned} Bel(v_j) &= 0 \\ Pl(v_j) &= 0 \end{aligned}$

   b. For each sub-hypercube $\bar{U}^l \in \Upsilon$:
   - Solve problem (45) on $\bar{U}^l$ and store $\nu^l_{\min}$ and $\nu^l_{\max}$.
   - If $\nu^l_{\max} \leq v_j$, then:

   $$Bel(v_j) = Bel(v_j) + BPA(\bar{U}^l)$$
   $$Pl(v_j) = Pl(v_j) + BPA(\bar{U}^l)$$

- Else, if $v^I_{\min} < v_j < v^I_{\max}$, partition $\bar{U}^I$ into two new sub-hypercubes $\bar{U}^{I1}$ and $\bar{U}^{I2}$. Remove $\bar{U}^I$ from $\Upsilon$ and add $\bar{U}^{I1}$ and $\bar{U}^{I2}$.
- Repeat step b. until a termination condition is met, e.g. the maximum number of partitions has been reached or the current $\bar{U}^I$ corresponds to a single focal element and therefore cannot by further divided. Alternatively further subdivisions are also avoided if the BPA of $\bar{U}^I$ is lower than a certain threshold, which means that its contribution to the Belief and Plausibility curves would be negligible.

(Note that step b. is to be skipped if problem (45) has already been solved on $\bar{U}^I$ and the results already stored are used instead).

c. For each $\bar{U}^m = \left[ \bar{U}^m \in \Upsilon \mid v^m_{\min} < v_j < v^m_{\max} \right]$:

$$Pl(v_j) = Pl(v_j) + BPA(\bar{U}_m)$$

We report here only the curves for designs 1 and 5, for the case without contamination only. These two are the most relevant since they correspond to the upper and lower edge of the *deterministic* Pareto front (see Fig. 8a). The curves for the other three design points are qualitatively similar. Fig. 10a and Fig. 10b show design point 5, corresponding to the lower left part of the Pareto front, i.e. minimum mass/minimum deviation.

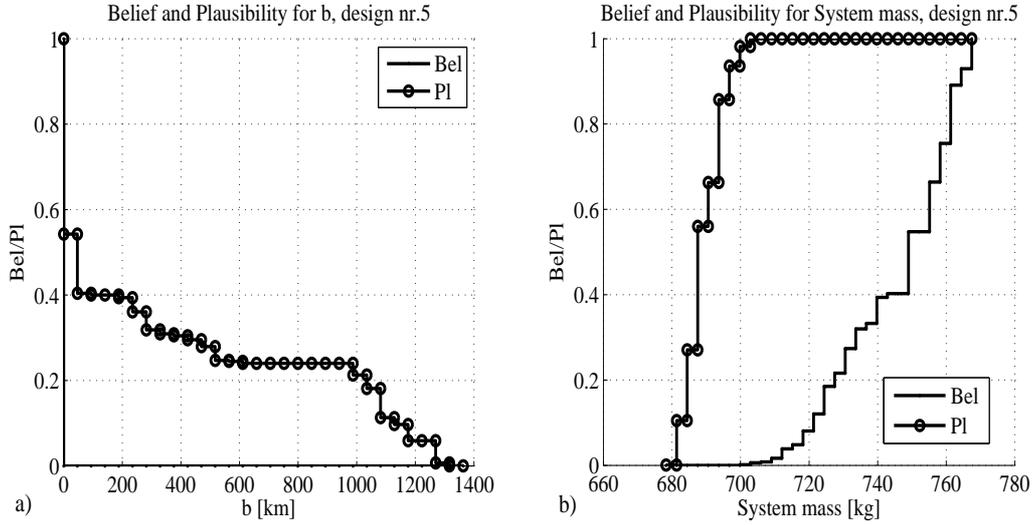

**Fig. 10** Belief/Plausibility curves for design 5: a) impact parameter *b* b) system mass

The deviation obtained is indeed very small, going from few tens of meters for *Bel*=1 to few thousands for *Pl*=0. At the same time, the curves of the system mass show that it cannot be lower than 680 *kg* but also will not exceed 765 *kg* even in the worst possible condition.

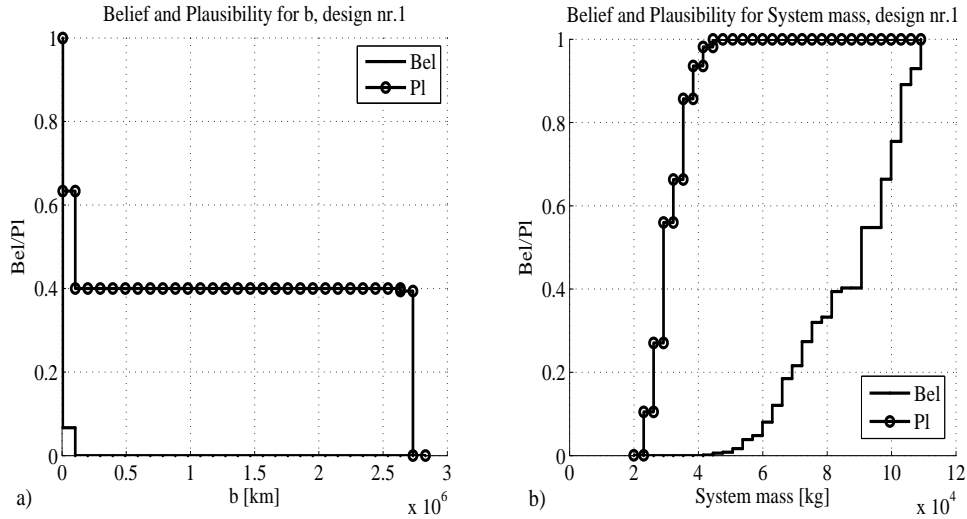

**Fig. 11** Belief/Plausibility curves for design 1: a) impact parameter *b* b) system mass

Similar observations are applicable to Fig. 11a and Fig. 11b, which consider the design point corresponding to maximum deviation and maximum system mass. In this case however, the difference between the condition with *Bel*=1 and *Pl*=0 is much wider: ~$10^6$-$10^4$ *km* for the impact parameter *b* and ~$1.09 \cdot 10^5$-$2 \cdot 10^4$ *kg* for the mass. This means that in the worst case, a successful deviation is still achievable, albeit with a small margin, but the total launch mass of the formation will be quite high. Note that, in the case of design 1, the performance values for worst and best conditions (*Bel*=1 and *Pl*=0) are coinciding with the values at upper edge of the *minmax* and *minmin* Pareto fronts respectively, as reported in Fig. 8a. This is explained by the fact that the design points corresponding to the upper edge of the *deterministic*, *minmin* and *minmax* curves are identical and correspond to the point with $n_{sc}=10$, $d_M=20$ *m* and $t_{warn}=8$ *years* as in Fig. 9. However, this is not the case in general (as already discussed in the previous section) and therefore for example the performance values for the Bel=1 and Pl=1 conditions for design points 2 to 4 will be different from the best case and worst case conditions defined by the *minmin* and *minmax* Pareto fronts.

It is interesting to observe that the Bel/Pl curves for *b* follows a *stepped* trend with three large variations while the mass' curves have a more gradual increase from 0 to 1. This possibly means that the impact parameter is mostly influenced by a single physical parameter rather than by a combination of many of them. In order to identify the most influent parameter, one can calculate the Belief and Plausibility curves for design point 1 with respect to each individual physical parameter while considering the remaining ones as constants with the values in Table 7. This analysis does not consider the coupling or interdependency of the parameter and therefore does not provide a complete picture of the impact of one uncertain parameter on the system performance. Nonetheless it gives a qualitative indication of the relative importance of the uncertain parameters. The results are shown in Fig. 12.

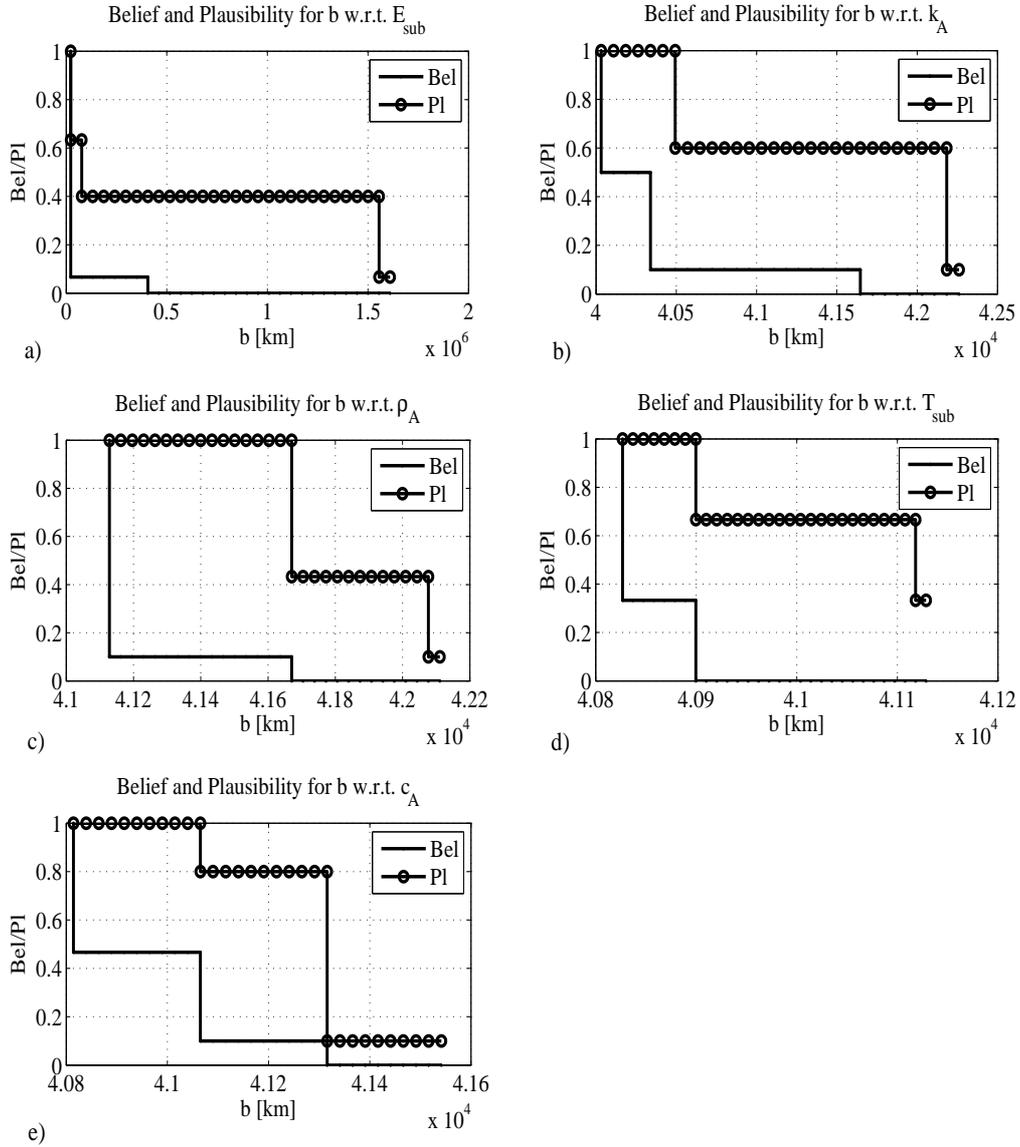

**Fig. 12** Belief/Plausibility curves for *b* w.r.t. $E_{sub}$

In Fig. 12a one sees that in the case of the sublimation enthalpy the difference in impact parameters between the points at *Bel*=1 and *Pl*=0 is much greater than in the four other cases (Fig. 12b to Fig. 12e).

This shows that the wide boundaries introduced on the enthalpy are a driving factor in determining the wide spreading between the best case and worst case impact parameter *b*. It also means that, with the current knowledge on the value of the sublimation enthalpy (see Table 5), a tight enclosure of the performance of the laser ablation system is not possible.

## 7. Conclusions

This work presented the combined orbital and system model for the multiobjective optimisation under uncertainties of the deflection of an asteroid with laser ablation. A fast and accurate analytical propagation of the low-thrust deflection action, though FPET, allowed for the fast computation of the Pareto set of optimal solutions for the asteroid deflection problem. The deterministic multiobjective optimisation showed that solar-pumped laser ablation can easily achieve considerable NEO deviations with a launch mass within current or near future technological capabilities. The uncertainty on some critical technologies and NEO physical characteristics were modelled and quantified through Evidence Theory. By including these uncertainties in the optimisation process, one can observe that in the worst case scenario the effectiveness of the whole concept is severely compromised. The analysis of the Belief and Plausibility curves has revealed that the sublimation enthalpy is the most critical uncertain parameter, due to its wide range of values which depend on asteroid type and also due to the disagreement of different sources. The optimisation approach under uncertainty proposed in this paper was demonstrated to be a useful tool to highlight the key knowledge areas which will require better investigation in the early phases of mission design. Furthermore, it provides a quantitative measure of which solutions should be adopted to be robust against current uncertainty.

### Acknowledgements

This research is partially supported through the ESA/ITI grant AO/1-5679/08/NL/CB.

### References

[1] McAdams, J.V., Dunham, D.W., Mosher, L.E., Ray, J.C., Antreasian, P.G., Helfrich, C.E., Miller, J.K.: Maneuver history for the NEAR mission—launch through Eros orbit insertion. In: Proceedings of the AIAA/AAS Astrodynamics Specialist Conference, AIAA-2000-4141 (2000)

[2] Glassmeier, K.H., Boehnhardt, H., Koschny, D., Kührt, E., Richter, I.: The rosetta mission: Flying towards the origin of the solar system. Space Sci. Rev. 128(1–4), 1–21 (2007)

[3] Rayman, M., Varghese, P., Lehman, D., Livesay, L.: Results from the deep space 1 technology validation mission. Acta Astronaut. 47(2), 475–487 (2000)

[4] Nakamura, A.M., Michel, P.: Asteroids and their collisional disruption. In: Lecture Notes in Physics, Small Bodies in Planetary Systems, pp. 1–27. Springer, Berlin (2009)

[5] Hampton, D., Baer, J., Huisjen, M., Varner, C., Delamere, A., Wellnitz, D., A'Hearn, M., Klaasen, K.: An overview of the instrument suite for the deep impact mission. Space Sci. Rev. 117(1–2), 43–93 (2005)

[6] Russell, C.T., Capaccioni, F., Coradini, A., De Sanctis, M.C., Feldman, W.C., Jaumann, R., Keller, H.U., McCord, T.B., McFadden, L.A., Mottola, S., S., Pieters, C.M., Prettyman, T.H., Raymond, C.A., Sykes, M.V., Smith, D.E. and Zuber, M.T.: Dawn mission to Vesta and Ceres. Earth Moon Planets 101(1–2), 65–91 (2007).

[7] Jet Propulsion Laboratory, Near Earth Object Program, http://neo.jpl.nasa.gov/neo/groups.html (2012). Accessed 7 April 2012.

[8] Project Icarus, MIT Press, Cambridge, MA (1968).

[9] Smith, P.L., Barrera, M.J. et al.: Deflecting a Near Earth Object with Today's Space Technology, AIAA planetary Defense Conference, Orange County, CA, AIAA Paper 2004-1447, Feb. (2004).

[10] McInnes, C.: Deflection of Near-Earth Asteroids by Kinetic Energy Impacts from Retrograde Orbits", Planetary and Space Science, Vol.52, No.7 (2004).

[11] Scheeres, D.J., Schweickart, R.L.: The Mechanics of Moving Asteroids, AIAA planetary Defense Conference, Orange County, CA, AIAA Paper 2004-1447, Feb. (2004).

[12] Lu, E.T., Love, S.G.: Gravitational Tractor for Towing Asteroids, Nature, Vol.438, Nov. (2005).

[13] Melosh, H.J., Nemchinov, I.V., Zetzer, I.I.: Non-nuclear Strategies for Deflecting Comets and Asteroids, Hazards Due to Comets and Asteroids, Univ. of Arizona (1994).

[14] Campbell, J.W., Phipps, C., Smalley, L., Reilly, J., Boccio, D.: The Impact Imperative: Laser Ablation for Deflecting Asteroids, Meteoroids, and Comets from Impacting the Earth, AIP Conference Proceedings, Vol. 664 (2003).

[15] Olds, J., Charania, A., Schaffer, M.G.: Multiple Mass Drivers as an Option for Asteroid Deflection Missions, 2007 Planetary Defense Conference, Washington, D.C., Paper 2007 S3-7, Mar. (2007).

[16] Spitale, J.N.: Asteroid Hazard Mitigation using the Yarkovsky Effect, Science, Vol. 296, No. 5565, (2002).

[17] Sanchez, J.P., Colombo, C., Vasile, M., Radice, G.: Multicriteria Comparison among several Mitigation Strategies for Dangerous Near-Earth Objects, Journal of Guidance, Control, and Dynamics, Vol. 32, No. 1, Jan-Feb (2009).

[18] Phipps, C.: Laser Deflection of NEO's, Report of the NASA Near-Earth-Object Interception Workshop, January 14-16 (1992).

19. Phipps, C.: Laser Deflection of near Earth asteroids and comet nuclei, In: Proceedings of the International conference on Lasers 96 (1997).
20. Phipps, C.: Can Lasers Play a Rôle in Planetary Defense?, AIP Conference Proceedings 1278, 502-508 (2010).
21. Park, S.Y., Mazanek, D.D.: Deflection of Earth-crossing asteroids/comets using rendezvous spacecraft and laser ablation, Journal of Astronautical Sciences, Vol. 53, No. 1, 21-37 (2005).Vasile, M., Maddock, C.A., Summerer, L.: Conceptual design of a multi-mirror system for asteroid deflection (2009).
22. Vasile M., Maddock C., Radice G., McInnes C.: NEO Deflection though a Multi-Mirror System, ESA Call for Proposals: Encounter 2029, Final Report for Ariadna Study Contract 08/4301, Technical officer: Summerer L. (2009).
23. Maddock, C.A., Vasile, M.: Design of optimal spacecraft-asteroids formation through a hybrid global optimization approach, International Journal of Intelligent Computing and Cybernetics, Vol. 1, No.2 (2008).
24. Gibbings, A., Vasile, M., Hopkins, J.M., Burns, D., Watson, I.: Exploring and Exploiting Asteroids with Laser Ablation, UK Space Conference, Warwick, UK, 4th-5th July (2011).
25. Gibbings, A., Vasile, M., Hopkins, J.M., Burns, D.: On testing Laser ablation processes for asteroid deflection, 2011 IAA Planetary Defence Conference, Bucharest, Romania, 9-12th May (2011).
26. Vasile, M. and Croisard, N.: Robust Preliminary Space Mission Design under Uncertainty, Computational Intelligence in Expensive Optimization Problems, 543-570, Springer (2010).
27. Zuiani, F., Vasile, M., Palmas, A., Avanzini, G.: Direct Transcription Of Low-Thrust Trajectories With Finite Trajectory Elements, Acta Astronautica (2011).
28. Vasile, M. and Colombo, C.: Optimal impact strategies for asteroid deflection, Journal of Guidance, Control, and Dynamics, Vol. 31, No. 4 (2008).
29. Colombo, C., Vasile, M. and Radice, G.: Semi-analytical solution for the optimal low-thrust deflection of Near-Earth Objects, Journal of Guidance, Control and Dynamics, Vol. 32, No. 3, 796-809 (2009).
30. Vasile, M. and Maddock, C.A.: On the deflection of asteroids with mirrors, Celestial Mechanics and Dynamical Astronomy, Vol. 107, No. 1, 265-284 (2010).
31. Battin, R.H.: An introduction to the mathematics and methods of astrodynamics (1999).
32. Palmas, A.: Approximations of low-thrust trajectory arcs by means of perturbative approaches, M.D. Thesis, Politecnico di Torino (2010).
33. Vasile, M., Maddock, C., and Radice, G.: Mirror formation control in the vicinity of an asteroid. In: AIAA/AAS Astrodynamics Specialist Conference and Exhibit, 18-21 August 2008, Honolulu, Hawaii (2008).
34. Kahle, R., Kuhrt, E., Hahn, G., Knollenberg, J.: Physical limits of solar collectors in deflecting Earth-threatening asteroids, Aeropace science and technology, Vol. 10, No. 3 (2006).
35. Wertz, J.R. and Larson, W.J.: Space mission analysis and design, Microcosm (1999).
36. Phipps, C., Bohn, W., Eckel, H.A., Horisawa, H., Lippert, T., Michaelis, M., Rezunkov, Y., Sasoh, A., Schall, W., Scharring, S. and others, Review: Laser-Ablation Propulsion, Journal of Propulsion and Power, Vol. 26, No.4, 609-637 (2010).
37. Klir, G.J., Smith, R.M.: On Measuring Uncertainty and Uncertainty-Based Information: Recent Developments, Annals of Mathematics and Artificial Intelligence, Vol. 32, No.1-4 (2001).
38. Britt, D.T., Yeomans, D., Housen, K. and Consolmagno, G.: Asteroid Density, Porosity, and Structure, Asteroids III, Vol.1, 485-500 (2002).
39. Pieters, C.M. and McFadden, L.A.: Meteorite and Asteroid Reflectance Spectroscopy: Clues to Early Solar System Processes, Annual Review of Earth and Planetary Sciences, Vol. 22, 457-497 (1994).
40. Price, S.D.: The surface properties of asteroids, Advances in Space Research, Vol. 33, No. 9, 1548-1557, Elsevier (2004).
41. Oberkampf, W., Helton, J.: Investigation of Evidence Theory in Engineering Applications, AIAA-Paper 2002-1569 (2002).
42. Vasile, M., Zuiani, F.: Multi-agent collaborative search: an agent-based memetic multi-objective optimization algorithm applied to space trajectory design, Proceedings of the Institution of Mechanical Engineers, Part G: Journal of Aerospace Engineering (2011).
43. Vasile, M., Minisci, E. and Locatelli, M.: An inflationary differential evolution algorithm for space trajectory optimization, IEEE Transactions on Evolutionary Computation, Vol. 15, No. 2, 267-281 (2011).
44. Vasile, M., Minisci, E., Zuiani, F., Komninou, E., Wijnands, Q.: Fast Evidence-based Systems Engineering, Paper IAC-11.D1.3.3, 62nd International Astronautical Congress, Cape Town, South Africa, 3rd-7th October (2011).